\documentclass[letterpaper, journal,twoside, twocolumn]{support/IEEEtran}
\usepackage[fleqn]{amsmath}
\usepackage{times}
\usepackage[pdftex]{graphicx}
\usepackage{subfigure}
\usepackage{amsmath,amssymb,amsopn,amstext,amsfonts}
\usepackage{cancel}
\usepackage[noadjust]{cite}
\usepackage{soul}
\usepackage{caption}
\usepackage{balance}
\usepackage{color}
\usepackage{mathtools}
\usepackage{bm}
\usepackage{diagbox}
\usepackage{float}
\usepackage{epstopdf}
\usepackage{url}
\usepackage{multirow}
\usepackage{tikz}
\usepackage{subeqnarray}
\usepackage{cases}
\usepackage{booktabs}
\usepackage[linkcolor=black,citecolor=black,urlcolor=black,colorlinks=true]{hyperref}
\usepackage{algorithm}
\usepackage[noend]{algpseudocode}
\newtheorem{myTheo}{Theorem}
\newtheorem{myDef}{Definition} 
\newtheorem{lemma}{Lemma} 

\newtheorem{remark}{Remark}

\newtheorem{assumption}{Assumption}
\newtheorem{example}{Example}
\soulregister\cite7
\soulregister\citep7
\soulregister\citet7
\soulregister\ref7
\soulregister\it7
\soulregister\pageref7

\graphicspath{{figures/}}
\DeclareGraphicsExtensions{.pdf,.png,.jpg,.eps}
\IEEEoverridecommandlockouts

\title{\LARGE \bf  Resilient   Output Consensus Control  of Heterogeneous Multi-agent Systems against Byzantine Attacks: A Twin Layer Approach}
\author{
  \vskip 1em
  {
  Xin Gong, \emph{Graduate Student Member, IEEE}, Yiwen Liang,
   Yukang Cui, \emph{Member, IEEE},
   Shi Liang,
  and Tingwen Huang, \emph{Fellow,~IEEE}
  }

  \thanks{
 This work was partially supported by the National Natural Science Foundation of China under Grant 61903258, Guangdong Basic and Applied Basic Research Foundation 2022A1515010234 and the Project of Department of Education of Guangdong Province 2022KTSCX105. 

X. Gong and S. Liang are with the Department of Mechanical Engineering, The University of Hong Kong, Pokfulam Road, Hong Kong (e-mail: {\tt\small gongxin,lswyy@connect.hku.hk}).

Y. Liang and Y. Cui are with the College of Mechatronics and Control Engineering, Shenzhen University, Shenzhen, 518060, China (e-mail: {\tt\small yw3liang,cuiyukang@gmail.com}).

T. Huang is with Texas A\&M University at Qatar, Doha, 23874, Qatar (e-mail: {\tt\small tingwen.huang@qatar.tamu.edu}).

  }
}

\begin{document}
  \maketitle
  \begin{abstract}
    This paper studies the problem of cooperative control of heterogeneous multi-agent systems (MASs) against Byzantine attacks. The agent affected by Byzantine attacks sends different wrong values to all neighbors while applying wrong input signals for itself, which is aggressive and difficult to be defended. Inspired by the concept of Digital Twin, a new hierarchical protocol equipped with a virtual twin layer (TL) is proposed, which decouples the above problems into the defense scheme against Byzantine edge attacks on the TL and the defense scheme against Byzantine node attacks on the cyber-physical layer (CPL). On the TL, we propose a resilient topology reconfiguration strategy by adding a minimum number of key edges to improve network resilience. It is strictly proved that the control strategy is sufficient to achieve asymptotic consensus in finite time with the topology on the TL satisfying strongly $(2f+1)$-robustness. On the CPL, decentralized chattering-free controllers are proposed to guarantee the resilient output consensus for the heterogeneous MASs against Byzantine node attacks. Moreover, the obtained controller shows exponential convergence. The effectiveness and practicality of the theoretical results are verified by numerical examples.

\end{abstract}
\begin{IEEEkeywords}
     Cooperative control, Heterogeneous MASs, Byzantine attacks, Resilient control
\end{IEEEkeywords}
\section{Introduction}
\IEEEPARstart The coordination issues of multi-agent systems (MASs) have attracted considerable attention in the robotics communities and the control societies due to its wide applications in the formation of unmanned aerial vehicles (UAVs) \cite{2017Time}, electrical power grids \cite{2017Distributed}, attitude alignment of satellites \cite{2020Robust}, aggregation behavior analysis of animals \cite{2004Consensus}, etc. It is well known that leader-following consensus can be reached by sufficient local neighboring relative interactions. As the state variables of each agent cannot always be available in practice, various output feedback-based control protocols have been developed for event-triggered MASs \cite{2015Event}, MASs with time delays \cite{2016A}, high-order nonlinear MASs \cite{2019Leader}, etc. It should be pointed out that these results commonly have an assumption that the networked system is homogeneous with all dynamics of coupled agents being identical. However, in real engineering applications, the networked system is usually heterogeneous. Considering this issue, the leader-following consensus of heterogeneous MASs has been studied in \cite{2018Consensus,2019LeaderS,2020A,luo2021leader}. In \cite{2018Consensus}, distributed consensus controllers were designed to guarantee the leader-following consensus with heterogeneous input saturation. The authors in \cite{2019LeaderS} investigated leader-following consensus of coupled heterogeneous harmonic oscillators that are one kind of second-order MASs based on relative position measurements. The leader-following consensus problem of heterogeneous MASs with nonlinear units and communication time-delay was studied in \cite{2020A} and \cite{luo2021leader}, respectively.

In practical applications, modern large-scale complex networked systems are vulnerable to breaking down when one or more nodes are compromised and become non-cooperative, which may be caused by malicious attacks on the communication network. In order to improve the safety of networks, the resilient consensus has been a vital research topic in the control system community in the presence of malicious attacks such as malicious attacks \cite{2020ResilientZ}, Byzantine attacks \cite{1982The}, denial of service attacks (DoS) \cite{2016ResilientY}, false-data injection (FDI) attacks \cite{2018State}, deception attacks \cite{2020Secure}, and replay attacks \cite{2020Event}. The resilient control of MASs aims to develop distributed consensus protocols that provide an admissible system performance in hostile situations despite the network misbehavior. In the first study of the resilience of consensus under Byzantine attacks in \cite{1982The}, the problem was described abstractly in terms of a group of Byzantine generals with their troops camped in different places around an enemy city. However, some of them may be traitors who will confuse others to reach an agreement, which are referred to as Byzantine agents and Byzantine attacks. Since this pioneering work of the analysis of Byzantine attacks, several important advances towards the resilient consensus have been made in the last decades. In \cite{1994Reaching}, a series of algorithms, called the Mean-Subsequence-Reduced (MSR) algorithms were developed to handle the approximate Byzantine consensus problem, in which the loyal nodes are able to achieve approximate agreement in the presence of $f$-total Byzantine faults in finite time. Then, authors in \cite{2012Low} put forward a continuous-time variation of the MSR algorithms, referred to as the Adversarial Robust Consensus Protocol (ARC-P), to achieve asymptotic consensus for Byzantine threat networks with constrained topology. In \cite{2013Resilient}, a new topological property, named network robustness, was first introduced to measure the ability of the developed local algorithms W-MSR to succeed. These concepts have been later extended and employed to analyze second-order MASs in \cite{2015Consensus}. Furthermore, \cite{2017Secure} investigated the consensus problem for nonlinear MASs suffering from attacks and communication delays. Besides the above works, researchers also investigate some other consensus-based settings like synchronization of linear time-invariant systems \cite{2017Resilient}, distributed observers \cite{2019Byzantine}, distributed optimization\cite{2017DistributedS}, and so on.

However, besides the aforementioned works \cite{2018Consensus,2019LeaderS,2020A,luo2021leader}, there are few results available for heterogeneous networks, which complicates the controller design since the agents with different dynamics are coupled through information exchange.  In this article, we propose a hierarchical resilient framework to achieve the consensus of heterogeneous multi-agent systems against Byzantine attacks. For addressing this issue, the main contributions can be summarized as follows:
\begin{enumerate}
\item Inspired by the concept of digital twin, we design a double-layer resilient control architecture, including twin layer (TL) and cyber-physical layer (CPL). As the TL is  deployed on the virtual space, the networked MASs are immune to Byzantine node attacks on the TL. Because of this feature of the TL, the resilient control strategy can be decomposed into  the defense against Byzantine edge attacks on TL and  the defense against Byzantine node attacks on CPL, as shown in Fig. \ref{image2}.
\item
We propose a resilient topology reconfiguration strategy by adding a minimum number of key edges to improve the network resilience of the TL. A new kind of control protocol is  developed based on the above topology to achieve the distributed
accurate estimation on each order of leader state.
\item A decentralized controller is designed on the CPL, which makes the tracking error exponentially attenuate to zero
even against  Byzantine attacks. Different from the existing work in \cite{2021ResilientY}, we manage to ensure both Byzantine nodes and normal nodes track the leader in the presence of Byzantine attacks. This implies that the network connectivity of the whole MASs is preserved thanks to our hierarchal approach. Thus, the cooperative control of agents could be stilled accomplished via our double-layer resilient control scheme, despite Byzantine attacks.
\end{enumerate}

\noindent\textbf{Notations:}
In this paper, { $\sigma_{{\rm min}}(S)$ and $\sigma(S)$ are the minimum singular value and the spectrum of matrix $S$,
respectively.} $|S|$ is the cardinality of a set $S$. $\Vert \cdot \Vert$ is the Euclidean norm of a vector. Define the sets of real numbers, positive real numbers, nonnegative integers, positive integers as $\mathbb{R}$, $\mathbb{R}_{> 0}$, $\mathbb{Z}_{\geq 0}$ and $\mathbb{Z}_{> 0}$. {The in-neighbor set and out-neighbor set denote the index sets denoted by $\mathcal{N}_i^{in}= \{i \in \textbf{I} \ | \ (z_j, z_i) \in \mathcal{E} \}$ and $\mathcal{N}_i^{out}= \{i \in \textbf{I} \ | \ (z_i, z_j) \in \mathcal{E} \}$, respectively.}
The Kronecker product is denoted by $\otimes$. $I_N \in \mathbb{R}^{N \times N}$ is the identity matrix. Denote $\textbf{I}[a,b]=\{a, a+1, \dots, b\}$  where $a<b$.
\section{Preliminaries}
\subsection{Robust Graph Theory}
For a group of agents, define a digraph $\mathcal{G } =(\mathcal{V}, \mathcal{E}, \mathcal{A} )$ with $\mathcal{V}=\{0, 1,\dots, N\}$, which indicates the edge set, $\mathcal{E} \subset \mathcal{V} \times \mathcal{V}$ indicates the edge set. The associated adjacency matrix is represented as $\mathcal{A} = [a_{ij}]$. An edge rooting from node $j$ and ending at node $i$ is represented by ($z_j$, $z_i$), meaning the information flow from node
$j$ to node $i$. The weight of edge
($z_j$, $z_i$) is $[a_{ij}]$, and $a_{ij} > 0$ when $(z_j, z_i) \in  \mathcal{E} $, otherwise $a_{ij} = 0$. $b_{i0}> 0 $ if there is an edge between the leader and the $i$th agent, otherwise $b_{i0} = 0$. Some useful definitions of network robustness are recalled first.

\begin{myDef}[r-reachable set]
Consider a graph $\mathcal{G } =(\mathcal{V}, \mathcal{E})$, a nonempty node subset $S \in \mathcal{V}$,  We say that $\mathcal{S}$ is an $r$-reachable set if at least one node exists in set $\mathcal{S}$ with at least $r$ neighbors outside set $\mathcal{S}$.
\end{myDef}\label{def41}

The notion of $r$-reachable is able be expressed as follows: Let $r \in \mathbb{Z}_{>0}$, then define $\mathcal{X}_S \subseteq \mathcal{S}$ to be the subset of nodes in $\mathcal{S}$, that is
\begin{equation*}
\mathcal{X}_S=\{i \in \mathcal{S}:|\mathcal{N}_i \backslash \mathcal{S}| \geq r  \}.
\end{equation*}

\begin{myDef}[strongly r-robust w.r.t. $\mathcal{X}$]
Consider a graph $\mathcal{G } =(\mathcal{V}, \mathcal{E})$ and a nonempty set $\mathcal{X} \subseteq \mathcal{V}$. $\mathcal{G }$ is called strongly $r$-robust w.r.t. $\mathcal{X}$ if every nonempty subset $\mathcal{Y} \subseteq \mathcal{V} \backslash \mathcal{X}$ is a $r$-reachable set.
\end{myDef}
\subsection{Nonsmooth Analysis}
Next, we review some essential results on nonsmooth analysis, which will be used in this paper to investigate the convergence of the distributed protocols considered.

Consider the  following system (possibly discontinuous)
\begin{equation}\label{eq1}
\dot{x}=f(x), ~\  x \in \mathbb{R}^n,
\end{equation}
where $f(x) : \mathbb{R}^n \rightarrow  \mathbb{R}^n$ is defined for every $x \in \mathbb{R}^n \backslash K$, where $K$ is a subset of $\mathbb{R}^n$ of measure zero.
There exists a constant $k$ such that $\Vert f(x) \Vert \leq k$.
We apprehend the  corresponding solution in the  sense of Filippov as the solution of an appropriate differential inclusion when the \eqref{eq1} has discontinuous right-hand
side.

\begin{myDef}[Filippov Solution]
A vector $x(t)$ is a solution of \eqref{eq1} for $t \in [t_a,t_b]$ if $x(t)$ is absolutely continuous on the time interval $[t_a,t_b]$.
\begin{equation}\label{eq2}
\dot{x} \in H[f](x),
\end{equation}
where $H[f](x) : \mathbb{R}^n \rightarrow 2^{\mathbb{R}_n}$ is defined as
\begin{equation}\label{eq3}
 H[f](x) \triangleq \underset{\alpha >0}{\bigcap} \ \underset{\delta\{K\}=0}{\bigcap} \hat{co}\{ f(A(x,\alpha) \backslash K )       \},
\end{equation}
where $\bigcap_{\delta\{K\}=0}$ denotes the intersection of all sets $K$ of measure zero, $A(x,\alpha)$ denotes the ball of radius $\alpha$ centered at $x$, $\hat{co}$ denotes the convex closure.

The tangent vector of Filippov's solution must lie in the convex closure of the vector field value in an area around the point of solution where the neighborhood becomes gradually smaller. It is important to note that the measurement zero set can be discarded under this definition.  If $f(x)$ is measurable and locally bounded, then the set-valued mapping $H[f](x)$ is upper semicontinuous, compact, convex and locally bounded, so (2) has Filippov solutions for each initial condition $x_0$.
\end{myDef}

\begin{myDef}[Clarke's Generalized Gradient \cite{clarke1990optimization}] Define $V(x)$ as a locally Lipschitz continuous function and define its Clarke's generalized gradient $\partial V(x)$  as
\begin{equation}\label{eq4}
\partial V(x) \triangleq co\{ \lim_{i \rightarrow \infty} \nabla V(x_i) | x_i \rightarrow x , x_i \notin \Phi_V \cup P
\},
\end{equation}
\end{myDef}
where $\nabla V$ denotes the  gradient, $x_i$ denotes a point which  converges to $x$ as $i$ tends to infinity, $\Phi_V$  is a set of Lebesgue measure zero
which contains all points without  $\nabla V$, and $P$ denotes a set of measure zero.
Then we  review the chain rule that allows to discriminate Lipschitz regular functions.

\begin{myTheo}[Chain Rule\cite{shevitz1994lyapunov}]
Let $x$ be a Filippov solution in \eqref{eq1} and $V(x(t))$ be a Lipschitz  regular function. Then $V$ is absolutely continuous, and
\begin{equation}
\frac{{\rm d}}{{\rm d} t}V(x(t)) \in \dot{\widetilde{V}}(x(t)),
\end{equation}
where the set-valued Lie derivative $\dot{\tilde{V}}(x(t))$ is
\begin{equation}
\dot{\tilde{V}}(x(t)) \triangleq \underset{\phi \in \partial V(x)}{\bigcap} \phi^T H[f](x).
\end{equation}
\end{myTheo}

Define the discontinuous “sign” function and the set-valued “SIGN” function as:
\begin{equation}
{\rm sign}(x)=
\begin{cases}
1, \hfill  {\rm if} \ x>0,\\
0, \hfill  {\rm if} \ x=0,   \\
-1, \hfill  {\rm if} \ x<0,
\end{cases}
\end{equation}
and
\begin{equation}
{\rm SIGN}(x)=
\begin{cases}
1, \hfill  {\rm if} \ x>0,\\
[-1, 1], \hfill  {\rm if} \ x=0,   \\
-1, \hfill  {\rm if} \ x<0.
\end{cases}
\end{equation}

\section{Problem Formulation}
\subsection{System Model and Related Assumptions} For a multi-agent system consisting of $N + 1$ agents, the agents assigned as followers are indexed by $\mathcal{F}=\{1,\ldots, N\}$ and the agent assigned as leader is indexed by $\mathcal{L}=\{0\}$. Then the dynamics of the followers are described as
\begin{equation}\label{eq6}
\begin{cases}
\dot{s}_i(t) = v_i(t),\\
\dot{v}_i(t)=\alpha_{s_i}s_i(t)+\alpha_{v_i}v_i(t)+u_i(t),\\
y_i(t)=s_i(t),
\end{cases}
\end{equation}
where $s_i \in \mathbb{R}^m$ , $v_i \in \mathbb{R}^m$, $u_i \in \mathbb{R}^m$ and $y_i \in \mathbb{R}^m$ are the position, velocity, control input and output of the follower,  respectively. Define $x_i = [s_i^T,v_i^T]^T $, $ i \in \mathcal{F} $. Then we have $\dot{x}_i(t) = A_i x_i(t)+B_i u_i(t)$ and $y_i(t)=C_i x_i(t)$.

 The dynamic of the leader is described as
 \begin{equation}\label{eq7}
\begin{cases}
\dot{s}_0(t) =  v_0(t),\\
\dot{v}_0(t) =  u_0(t),\\
y_0(t)=s_0(t),
\end{cases}
\end{equation}
where $s_0   \in \mathbb{R}^q$ , $v_0 \in \mathbb{R}^q$, $u_0 \in \mathbb{R}^q$ and $y_0   \in \mathbb{R}^q$ are the position, velocity, control input and output  of the leader, respectively.
Define $x_0=[s_0^T, v_0^T]^T$.
Then we have $\dot{x}_0(t)= S x_0(t)+D u_0(t)$ and $y_0(t)=R x_0(t)$.

\subsection{Related Assumptions}
Then we next make some assumptions.
\begin{assumption}\label{assumption_1}
The control input $u_0$ is upper bounded by a positive constant $u_{\rm max}$, which means ${\rm sup}_{t \geq 0} \Vert  u_0(t) \Vert _{\infty} \leq u_{\rm max}$.
\end{assumption}
\begin{assumption}\label{assumption_2}
The pair ($S$, $R$) is detectable.
\end{assumption}
\begin{assumption}\label{assumption_3}
For any $\lambda \in \sigma(S)$, it holds that\\
\begin{equation*}
{\rm rank} \bigg( \left[
  \begin{array}{ccc}
 A_i-\lambda I &  B_i   \\
C_i &  0   \\
  \end{array}
  \right]\bigg)=n_i+q.
\end{equation*}

\end{assumption}

\begin{lemma} (\cite{huang2004nonlinear})\label{lemma_3}
Under Assumptions $1$-$3$, for each agent $i$, there exist some matrices $\Gamma_i$ and $\Pi_i$ that satisfy the following regulator equations:
\begin{equation}\label{eq8}
\begin{cases}
A_i \Pi_i +B_i\Gamma_i= \Pi_i S, \\
C_i \Pi_i-R=0.
\end{cases}
\end{equation}
\end{lemma}

{

In this paper, only a fraction of agents is pinned to the leader.
}
For the sake of illustration, we take the agents where the leader can be directly observed as the pinned followers. The set of these pinned followers is described as set $\mathcal{V}_p$. On the other hand, the remaining agents are regarded as non-pinned followers and  collected in set $\mathcal{V}_{np}$.

Define the following local output consensus error:
\begin{equation}\label{eq9}
e_i(t)=y_i(t)-y_0(t).
\end{equation}

Based on the above settings, the heterogeneous MASs in (9) are said to achieve cooperative Control if $\lim_{t \to \infty} e_i(t) = 0, \forall i \in \mathcal{F}$.

\subsection{Byzantine Attack Model}\label{2B}

\begin{figure}
  \centering
  \includegraphics[height=7.4cm, width=8.5cm]{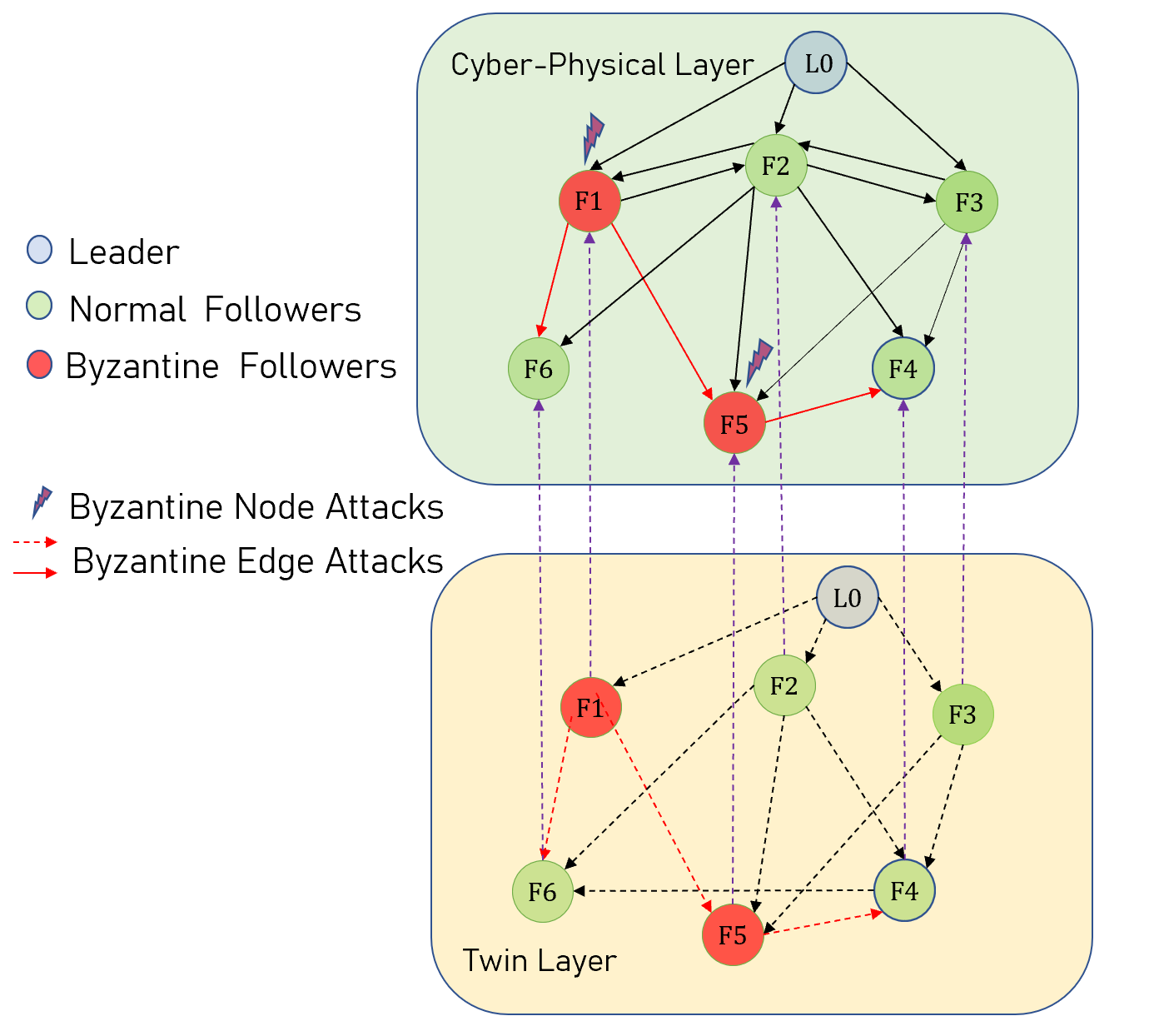}
  \caption{Distributed attack-resilient MASs against Composite Attacks: A two-layer framework.}\label{image2}
\end{figure}

\subsubsection{Edge Attacks}
We consider a subset $\mathcal{B} \subseteq \mathcal{V}$ of the nodes in the network to be adversarial. We assume that the nodes are completely aware of the network topology and the system dynamics of each agent. Let us denote the agents as Byzantine agents. Then, the normal agents are collected in the set $\mathcal{K}=\{1,2,\dots,k\}$, where $k$ is the number of normal agents in the network.

\begin{assumption}[f-local Attack]\label{assumption_5}
{
There exist at most $f$  Byzantine agents in the in-neighborhood of each agent, that is, $|\mathcal{B}\cap \mathcal{N}^{in} _i|\leq f , \ \forall i \in \mathcal{F}$.}
\end{assumption}

Here, we mainly consider the $f$-local attack model to deal with lots of Byzantine nodes in the network. It is reasonable to assume that there exist at most $f$ adjacent Byzantine nodes. Otherwise, it will be too pessimistic to protect the network.

\begin{remark}
The Byzantine nodes can completely understand the network topology and the system
dynamics. Different from general malicious nodes in \cite{2020ResilientZ}, the Byzantine nodes herein can send arbitrary and different false data to different in-neighbors and  cooperate with other Byzantine nodes at any time.
$\hfill \square$
\end{remark}

\subsubsection{Node Attacks and the associated TL solution}

The input signals of all Byzantine nodes on the CPL could be falsified via Byzantine node attacks. Under the influence of the Byzantine node attacks, the state of each Byzantine agent will be subverted as
 \begin{equation*}
   \bar{x}_i(t)=x_i+\psi_i(t),\ \forall i\in  \mathcal{B},
 \end{equation*}
 where $\psi_i(t)$ represents an unknown, time-varying, and potentially bounded signal satisfying Assumption \ref{assumption_6}.

 \begin{assumption}\label{assumption_6}
   The  magnitude of the Byzantine node attacks on each agent is bounded by  ${\kappa_i}>0$, that is, $\sup_{t\in \mathbb{R}_{\geq 0}}   |\psi_i(t)| \leq {\kappa_i}$, $\forall i \in \mathcal{F}$.
 \end{assumption}

Note that the $i$-index set of Assumption \ref{assumption_5} includes the Byzantine agents, apart from normal ones. This means, not only do we try to ensure that the normal agents are free from the compromised information from Byzantine edge attacks, but also we consider correcting the performance of Byzantine agents. As shown in Figure \ref{image2}, a supervising layer, named TL, provides another control signals to fight against the Byzantine node attacks. With the reference signals of TL, the real input signal of each agent consists of two parts:
 \begin{equation*}
   \bar{u}_i(t)=m_i(t){u}_i(t)+\psi_i(t),\ \forall i\in  \mathcal{F},
 \end{equation*}
 where $m_i(t)$ denotes a nonzero bounded factor  caused
by actuator faults  and
 $u_i(t)$ is the input {signal (correction signal)} to be designed later, which is concerned about the virtual states on the TL.

 \subsection{Problem Formulation}

Based on the above discussions, the resilient cooperative control problem of MASs, against Byzantine attacks will be summarized as follows:

%

\vspace{0.2cm}

\noindent \textbf{Problem RC3HPB} (Resilient Cooperative  Control of Heterogeneous MASs against Byzantine attacks):
Under Assumptions \ref{assumption_1}$\sim$\ref{assumption_6}, consider the agents in \eqref{eq6} subject to Byzantine attacks defined in Section \ref{2B}, design distributed protocols $u_i$ such that the global error $e_i$ in (\ref{eq9}) exponentially converges to zero.
$\hfill \hfill \square $

\section{Main Results}

Motivated by the recently sprung-up digital twin technology, a double-layer resilient control scheme is investigated in this section, as shown in Fig. \ref{image2}.
Owing to the introduction of TL, the resilient control scheme against BAs can be decoupled into the defense against Byzantine edge attacks on the TL and the defense against potentially  Byzantine node attacks on the CPL. More specifically, this hierarchal control scheme solves the \textbf{Problem RC3HPB} by employing a TL with an edge-adding strategy against Byzantine edge attacks and a decentralized adaptive controller on the CPL against potentially  Byzantine node attacks.

\begin{figure}
  \centering
  \includegraphics[height=5cm, width=8.9cm]{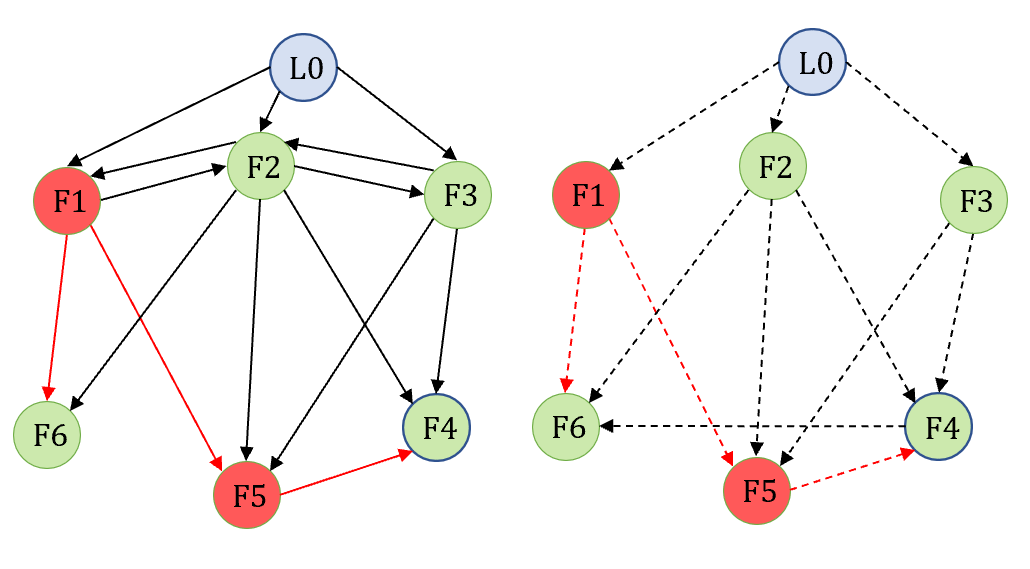}
  \caption{Topologies of the CPL and the TL.}\label{image3}
\end{figure}

%
%

\subsection{Distributed Resilient Estimation on the TL against Byzantine edge attacks}\label{IV_A}
As shown in Fig. \ref{image2}, there exists a TL governor to modify the network topology by adding a minimum number of key edges to improve the network resilience of the TL topology.
\subsubsection{Resilient Twins Layer Design: A Minimum Edge Number Approach}
Herein, we propose a heuristic algorithm to figure out the exact topology of TL, as shown in Algorithm 1. An illustrative example for Algorithm 1 is shown in
Fig. $\ref{image3}$, whose detailed process can be found in Example 1. Notice that $\mathcal{G}$ and $\mathcal{G}_T$ are not necessarily the same, since the TL has certain programmable flexibility in topology. Notice that the network topology $\mathcal{G}_T$ except the adding edges $(v_4,v_6)$ on the TL is a subset of the unweighed network topology $\bar{\mathcal{G}}=(\mathcal{V}, \mathcal{E})$ on the CPL. The adding edges are considered reliable since they are not affected by attackers.
Denote the in-neighbor and out-neighbor sets of node $i$ on the TL as $\mathcal{N}_{T,i}^{in}$ and $\mathcal{N}_{T,i}^{out}$, respectively. Now
we understand the algorithm through an example.

\begin{algorithm}
\caption{TL Topology Configuration Algorithm: A Minimum Edge Number Approach}
{\bf Input}: The topology $\mathcal{G}(\mathcal{V}, \mathcal{E}, \mathcal{A})$ and the parameter $f$.
\\
{\bf Output}: The TL topology $\mathcal{G}_T(\mathcal{V}_T, \mathcal{E}_T)$.
\begin{algorithmic}[1]
   \State Initialize $\mathcal{V}$ and $\mathcal{E}$  as $\varnothing$. Initialize the sets $\mathcal{M}_a = \varnothing$ and $\mathcal{M}_b$ = $\mathcal{F} \cup \mathcal{L}$.
   \State Add the leader node into $\mathcal{M}_a$. Update $\mathcal{M}_b=\mathcal{M}_b \backslash \mathcal{T}$.
   \While {$\mathcal{M}_a \neq \varnothing$}
   \State Denote  the nodes pinned to  $\mathcal{M}_a$ as $\mathcal{F}_u$.
   \While {$\mathcal{F}_u \ne \varnothing$}
   \For {$i \in \mathcal{F}_u$}
   \If {Node $i$ is pinned to the leader or at least $(2f+1)$ nodes in $\mathcal{M}_a$}
   \State Update $\mathcal{M}_a=\mathcal{M}_a \cup \{i\}$.
   \State Update $\mathcal{F}_u=\mathcal{F}_u \backslash \{i\}$.
   \State Update
   $\mathcal{V}_T$ as the set whose nodes are in $\mathcal{M}_a$.
   \State Update $\mathcal{E}_T=\mathcal{E}_T \cup \mathcal{E}_i$.
   \State Update $\mathcal{G}_T=(\mathcal{V}_T, \mathcal{E}_T)$.
   \Else
   \State Add an edge $(z_j, z_i)$ from node $j \in \mathcal{M}_a$ to node $i$ and node $j$ pins to the minimal out-neighbors.
   \EndIf
   \EndFor
   \EndWhile
   \EndWhile
 \State \Return The TL topology $\mathcal{G}_T(\mathcal{V}_T, \mathcal{E}_T)$.
\end{algorithmic}
\end{algorithm}

\begin{example} Consider the network topology in Fig. \ref{image2}:
\end{example}
{
\begin{enumerate}
\item Add the leader into $\mathcal{M}_a$. Update $\mathcal{M}_a={0}$ and $\mathcal{M}_b=\mathcal{F}$.
\item Collect the agents pinned to  $\mathcal{M}_a$ as $\mathcal{F}_u=\{1, 2, 3\}$.
Since the nodes in $\mathcal{F}_u$ is pinned to the leader node, we update $\mathcal{M}_a=\{0, 1, 2, 3\}$, $\mathcal{F}_u=\{4, 5, 6\}$, $\mathcal{V}_T=\{z_0, z_1, z_2, z_3\}$ and $\mathcal{E}_T=\{(z_0, z_1), (z_0, z_2), (z_0, z_3)\}$.
\item Collect the agents pinned to $\mathcal{M}_a$ as $\mathcal{F}_u=\{4, 5, 6\}$.\\ As only node 5 pins to at least $(2f+1)$ nodes in $\mathcal{M}_a$, we update $\mathcal{M}_a=\{0, 1, 2, 3, 5\}$, $\mathcal{F}_u=\{4, 6\}$,\\ $\mathcal{V}_T=  \{z_0, z_1, z_2, z_3,z_5\}$ and $\mathcal{E}_T=\{(z_0, z_1), (z_0, z_2),\\ (z_0, z_3), (z_1, z_5), (z_2, z_5), (z_3, z_5)\}$.

\item Collect the agents pinned to  $\mathcal{M}_a$ as $\mathcal{F}_u=\{4, 6\}$.\\ Since only node 4 pins to at least $(2f+1)$ nodes  in  $\mathcal{M}_a$, we update $\mathcal{M}_a=\{0, 1, 2, 3, 4,  5\}$, $\mathcal{F}_u=\{ 6\}$, $\mathcal{V}_T=\{z_0, z_1, z_2, z_3, z_4, z_5\}$ and $\mathcal{E}_T=\{(z_0, z_1), (z_0, z_2), (z_0, z_3), (z_1, z_5), (z_2, z_5), (z_3, z_5),\\ (z_2, z_4), (z_3, z_4), (z_5, z_4)\}$.
\item Collect the agents pinned to $\mathcal{M}_a$ as $\mathcal{F}_u=\{6\}$. As none of node  pins to the leader or at least $(2f+1)$ nodes in $\mathcal{M}_a$, the edge $(z_4, z_6)$ should be added. Then we can find that node 4 pins to at least $(2f+1)$ nodes in $\mathcal{M}_a$, we update $\mathcal{M}_a = \{0, 1, 2, 3, 4,  5, 6\}$, $\mathcal{F}_u=\{ 6\}$, $\mathcal{V}_T=\{z_0, z_1, z_2, z_3, z_4, z_5, z_6\}$ and $\mathcal{E}_T=\{(z_0, z_1), (z_0, z_2), (z_0, z_3), (z_1, z_5), (z_2, z_5), (z_3, z_5),\\ (z_2, z_4), (z_3, z_4), (z_5, z_4), (z_1, z_6), (z_2, z_6), (z_4, z_6)\}$.
\end{enumerate}
}



\subsubsection{MSR algorithm on the TL}
Then we use the MSR algorithm \cite{1994Reaching} to process the virtual state $z_i(t)$ on the TL. At any time $t$, each agent  always makes updates as below:
\begin{enumerate}
\item Collect the status of all neighbor agents (except the leader if $i \in \mathcal{V}_p$) in a list $\Delta_{i,s}(t)$, $s=\{1, 2 \}$.

\item The agents in $\Delta_{i,s}(t)$ are divided into $\overline{\Delta}_{i,s}(t)$ and $\underline{\Delta}_{i,s}(t)$ as follows :
\begin{equation*}
\begin{split}
&\overline{\Delta}_{i,s}(t)=\{j \in \Delta_i:z_{j,s}(t) > z_{i,s}(t)\},   \\
&\underline{\Delta}_{i,s}(t)=\{j \in \Delta_{i,s}:z_{j,s}(t) < z_{i,s}(t)\},
\end{split}
\end{equation*}

Remove $f$ largest state values in $\overline{\Delta}_{i,s}(t)$ that are greater than $z_{i,s}(t)$. Remove all values  if the number of agents in $\overline{\Delta}_{i,s}(t)$ is less than $f$.

\item Similarly, remove $f$ smallest state values in $\underline{\Delta}_{i,s}(t)$ that are lower than $z_{i,s}(t)$. Remove all values  if the number of agents in $\underline{\Delta}_{i,s}(t)$ is less than $f$.

\item Denote $\Omega_{i,s} (t)$, termed as an admitting set, as the collection of agents whose values are retained after (2) and (3).
\end{enumerate}

Then we formulate a distributed TL to achieve the resilient control of MASs:

\begin{equation}\label{eq16}
\begin{split}
\begin{cases}
\dot{z}_{i,1}=&z_{i,2}-c_1(a_{ij}\sum_{j \in \Omega_{i,1}(t)}{\rm sign}(z_{i,1}-z_{j,1})\\
&+b_{i0}{\rm sign}(z_{i,1}-z_{0,1})),  \\
\dot{z}_{i,2}=&-c_2(a_{ij}\sum_{j \in \Omega_{i,2}(t)}{\rm sign}(z_{i,2}-z_{j,2})\\
&+b_{i0}{\rm sign}(z_{i,2}-z_{0,2})),
\end{cases}
\end{split}
\end{equation}
where the virtual state ${z}_{i,j}$ denotes the $j$th   element of the $i$th agent's estimation on the leader, $c$ denotes a positive gain and $z_0$ denotes the state of leader that means $z_0=x_0$.

We then give a sufficient condition such that the distributed asymptotical estimation on the TL can be
achieved against the persistent Byzantine edge attacks:
\begin{myTheo}\label{theorem_2}
Consider a TL in (\ref{eq16})  under Assumptions \ref{assumption_1}$\sim$\ref{assumption_5} and  the $\mathcal{G} =(\mathcal{V}, \mathcal{E})$  is strongly $(2f+1)$-robustness w.r.t. $\mathcal{V}_p$. The estimation error on
the TL layer under Byzantine  attacks converges into zero asymptotically, if the conditions  $c_1 > 0$ and $c_2 > u_{\rm max}$ holds.
\end{myTheo}

\textbf{Proof.}
Consider the sets $I_s^M$ and $I_s^m$  as
\begin{equation*}
\begin{split}
&I_s^M=\{i \in \mathcal{F}:  z_{i,s}(t) > z_{0,s}(t) \},   \\
&I_s^m=\{i \in \mathcal{F}: z_{i,s}(t) < z_{0,s}(t) \}.
\end{split}
\end{equation*}
The nonsmooth Lyapunov function is introduced as
\begin{equation}\label{eq69}
V_s=V_s^M+V_s^m
\end{equation}
composed of terms
\begin{equation*}
V_s^M=\sum_{i \in I_s^M}(z_{i,s}-z_{0,s}),~
V_s^m=\sum_{i \in I_s^m}(z_{0,s}-z_{i,s}),
\end{equation*}
where  $V_s^M$ denotes the distance  between the followers with the largest state values and the leader and $V_s^m$ denotes the distance  between the followers with the smallest state values and the leader and $z_{0,s}$ denotes the $s$th element of the leader.

Next, we will mainly  analyze  the first term of \eqref{eq69}, since the term $V_p^m$ can be analyzed similarly.

Since $V_p^M$ is a Lipschitz regular function, we can use the chain rule given in Theorem 1 to obtain the set-valued Lie derivative as
\begin{equation}
\dot{\tilde{V}}_2^M(z_2)=\underset{\phi_2 \in \partial V_2^M(z_2)}{\bigcap} \phi_2^T H[f](z_2),
\end{equation}
where the generalized gradient $\partial V_2^M(z_2)$ and collective maps $H[f](z_2)$ are described next.
The generalized gradient can be expressed in the following form
\begin{flalign}\label{eq50}
\partial V_2^M(z_2)\!&\subseteq\! \sum_{i \in I_2^M} \!\! \partial z_{i,2}^M(z_2)\!
\\
&=\![0,\dots, Z_{k_1},\dots,Z_{k_q},\!-\!\sum_{i \in I_2^M}Z_i, \dots, 0]^T,&
\end{flalign}
where $Z_i= {\rm SIGN}(z_{i,2}-z_{0,2})$ and the indices $\{k_1,\dots,k_q\}$  represent the agents in the set $I_2^M$.

The composition of the $H[f](z_2)$ is
\begin{equation}
H[f](z_2)\subseteq [H[f_0](z_2),\dots,H[f_{N}](z_2)]^T,
\end{equation}
where terms $H[f_i](z_2)$ are defined as
\begin{align}
H[f_i](z_2) &= H \Big[  -c_2\sum_{j \in \Omega_{i,2}(t)} {\rm sign}(z_{i,2}-z_{j,2}) \Big]\notag\\
& \subseteq -c_2\sum_{j \in \Omega_{i,2}(t)}H[{\rm sign}(z_{i,2}-z_{j,2})]\notag\\
&=-c_2\sum_{j \in \Omega_{i,2}(t)} {\rm SIGN}(z_{i,2}-z_{j,2}),~\ i \in \mathcal{F}.
\end{align}
When the agent is a leader,
\begin{equation}
H[f_0](z_2)=u_{0}.
\end{equation}
Thus, we have
\begin{equation}\label{eq40}
\phi_2^T H[f](z_2) \subseteq \sum_{i= 0}^N \phi_{i,2} H[f_i](z_2).
\end{equation}

According to the definition of the set $I_s^M$, there exists at least a follower $i \in I_s^M$ whose status value is larger than that of the leader.
 In this case, we can calculate the function ${\rm SIGN}(z_{i,2}-z_{0,2})$ as
\begin{equation*}
{\rm SIGN}(z_{i,2}-z_{0,2})= \{1\}, ~ i \in I_2^M.
\end{equation*}
Thus, we obtain the following for the $\partial V_2^M$ :
\begin{align}
\partial V_2^M = [0, \dots, 1, \dots, 1, \dots, -|I_2^M|, \dots, 0]^T.
\end{align}
Then, \eqref{eq40} is further developed as
\begin{align}
&\sum_{i= 0}^N \phi_{i,2} H[f_i](z_2)
\\=&\sum_{i \in I_2^M} \Big(-c_2 ~ (a_{ij}\sum_{j \in \Omega_{i,2}(t)}{\rm SIGN}(z_{i,2}-z_{j,2})\\
&+b_{i0}{\rm SIGN}(z_{i,2}-z_{0,2}))-u_{0} \Big).
\end{align}

Since the neighbors of an agent are composed of the leader and followers, it yields
\begin{equation*}
\begin{split}
&c_2\sum_{j \in \mathcal{N}_i } {\rm SIGN}(z_{i,2}-z_{j,2})\\
=&c_2\sum_{j \in \mathcal{N}_i \cap \mathcal{L}} {\rm SIGN}(z_{i,2}-z_{j,2})+c_2\sum_{j \in \mathcal{N}_i \cap \mathcal{F}} {\rm SIGN}(z_{i,2}-z_{j,2})
\end{split}
\end{equation*}
According to \eqref{eq16}, the set-valued Lie derivative of $V_2^M$ can be calculated as
\begin{align}\label{eq70}
\dot{\tilde{V}}_2^M(z_2) =& \sum_{i \in I_2^M}(\dot{z}_{i,2}-\dot{z}_{0,2})\notag\\
=&\sum_{i \in I_2^M}\Big(-c_2 a_{ij}\sum_{j \in \Omega_{i,2}}{\rm SIGN}(z_{i,2}-z_{j,2})\notag\\
&-c_2 b_{i0}{\rm SIGN}(z_{i,2}-z_{j,2})-u_{0}\Big)\notag\\
\leq&-c_2|\mathcal{V}_p|-c_2|\mathcal{V}_{np}|+u_{\rm max}|\mathcal{F}|\notag\\
\leq&-(c_2-u_{\rm max})|\mathcal{F}|.
\end{align}

We consider the worst case, that is, there exist $f$ Byzantine agents  and only $f+1$ normal agents in the neighborhood. Thus, we can obtain that $-c_2\sum_{j \in \mathcal{N}_i \cap \mathcal{F}}{\rm sign}(z_{i,2}-z_{j,2})=-c_2$. Then, the set-valued Lie derivative can be proved as
\begin{equation}
-(c_2-u_{\rm max})|\mathcal{F}| < 0.
\end{equation}

Similarly, the Lie derivative of $V_1^M$ can be written as
\begin{flalign}
\dot{\tilde{V}}_1^M =& \sum_{i \in I_1^M}(\dot{z}_{i,1}-\dot{z}_{0,1})\notag&\\
=&\sum_{i \in I_1^M}\Big(z_{i,2}-z_{0,2}-c_1\sum_{j \in \mathcal{N}_i \cap \mathcal{F}}{\rm SIGN}(z_{i,1}-z_{j,1})\notag&\\
&-c_1\sum_{j \in \mathcal{N}_i \cap \mathcal{L}}{\rm SIGN}(z_{i,1}-z_{j,1})\Big).
\end{flalign}

Notice that Eq.\eqref{eq70} holds, which leads to $V_2^M=0$, that is, $z_{i,2}-z_{0,2}=0$. Thus, we have
\begin{align}
\dot{\tilde{V}}_1^M(z_1) &\leq -c_1|\mathcal{V}_p|-c_1|\mathcal{V}_{np}|\notag\\
&\leq -c_1|\mathcal{F}|< 0
\end{align}

Case 2: In this situation, the set $I_2^M$ is composed of  agents that share the same values as $z_{0,2}$. Inequality still hosts  when there exists at least
a follower in the set $I_2^M$.
The \eqref{eq50} can be written  as
\begin{align}
\partial V_2^M(z_2) &\subseteq \sum_{i \in I_2^M} \partial V_{i,2}^M(z_2)\notag\\
&=[0, \dots, \bar{Z},\dots,\bar{Z},-\sum_{i \in I_2^M}\bar{Z}, \dots, 0]^T,
\end{align}
where $\bar{Z} = {\rm SIGN}(z_{i,2}-z_{0,2})$.
Recalling \eqref{eq40}, we analyze the following dot product
\begin{flalign}\label{eq53}
&\sum_{i= 0}^N \phi_{i,2} H[f_i](z_2) \notag\\
=&\sum_{i \in I_2^M} \phi_{i,2}\Big(-c_2 \sum_{j \in \mathcal{N}_i}{\rm SIGN}(z_{i,2}-z_{j,2}) \Big)+\phi_{0,2} u_{0}.
\end{flalign}
Considering the  variation $[-u_{\rm max},  u_{\rm max}]$ for the control input of the leader and the fact that ${\rm SIGN}(z_{i,2}-z_{j,2}) \subseteq [-1, 1]$, the following result holds:
\begin{flalign}
&\sum_{i= 0}^N \phi_{i,2} H[f_i](z_2) \notag&\\
\subseteq &  ~ \phi_{0,2}     u_{0}+\! \! \! \sum_{i \in I_2^M }\! \!  \phi_{i,2} \Big(\! -c_2 \sum_{j \in \mathcal{N}_i } [-1, 1]+[-u_{\rm max},u_{\rm max}]\Big)\notag&\\ \subseteq&   ~ \phi_{0,2}   [-u_{\rm max}, u_{\rm max}]\notag&\\
&+\sum_{i \in I_2^M } \phi_{i,2} [-c_2|\mathcal{N}_i|-u_{\rm max},  c_2|\mathcal{N}_i|+u_{\rm max}].&
\end{flalign}

From the above results, we can conclude
\begin{equation}
\dot{\tilde{V}}_2^M(z_2)=\{0 \}.
\end{equation}

A similar analysis was performed for the $V_s^m$, replacing the set $I_s^M$ with the set $I_s^m$. Then, we can obtain the same bound as
\begin{equation*}
\begin{split}
\dot{\tilde{V}}_1^m(z_1) &\leq -c_1|\mathcal{F}|\\
\dot{\tilde{V}}_2^m(z_2) &\leq -(c_2-u_{\rm max})|\mathcal{F}|.
\end{split}
\end{equation*}

Recalling the fact that the time derivative $\frac{{\rm d}}{{\rm d} t} z_2^M(z_2) \in \dot{\tilde{V}}_2^M(z_2)$, the bound is as follows
\begin{equation}
\frac{\rm{d}}{\rm{d} t} V_2^M(z_2) \leq -(c_2-u_{\rm max})|\mathcal{F}|.
\end{equation}

Then,  since the  Lie derivatives of both  $V_2^M$ and $V_2^m$ upper bounded by $-(c_2-u_{\rm max})|\mathcal{F}|$, the bound on the derivative $\frac{{\rm d}}{{\rm d} t} V_2(z_2)$ can  be calculated as
\begin{equation}
\frac{\rm{d}}{\rm{d} t}V_2(z_2) \leq -2(c_2-u_{\rm max})|\mathcal{F}|,
\end{equation}
if there exists one follower that has not tracked the leader. In the meantime, we can conclude that when all followers track the leader, the derivative ${\rm d} V_2/{\rm d}t$
is equal to zero. For $V_1$, we can obtain the analogical analysis that $\frac{{\rm d}}{{\rm d} t}V_1(z_1)\leq -2c_1|\mathcal{F}|$.

This completes the proof.
$\hfill \blacksquare$
\begin{remark}
  {
  Different from the existing work \cite{3xiao2021distributed}, we consider the TL that can encounter Byzantine edge attacks, rather than the completely secure layer.
  Compared with the traditional distributed observer \cite{3xiao2021distributed}, the TL has stronger confidentiality and higher security, which can be deployed in the cloud. Also, since TL has certain programmability in topology, the topology on the TL $\mathcal{G}_T$ and the topology on the CPL $\mathcal{G}$ are not necessarily the same, as shown in Fig. \ref{image3}.
  $\hfill \square$
  }
  \end{remark}

\begin{myTheo}\label{theorem_3}
Consider the MASs satisfying Theorem \ref{theorem_2}. Then, the cooperative consensus can be achieved in a finite-time $T_s$ which is upper bounded by
\begin{flalign}
T_s=&\frac{1}{\chi_1}{\rm max}\{V_1^M(z_1(t_1)),V_1^m(z_1(t_1))\}\notag\\
&+\frac{1}{\chi_2}{\rm max}\{V_2^M(z_2(0)),V_2^m(z_2(0))\},
\end{flalign}
where $\chi_1=c_1|\mathcal{F}|$ and $\chi_2=(c_2-u_{\rm max})|\mathcal{F}|$.
\end{myTheo}
\textbf{Proof.}
Let us compute the $V_1^l(z_1(t_2))$ and $V_2^l(z_2(t_1))$ as follows:
\begin{flalign}
V_2^l(z_2(t_1))&=V_2^l(z_2(0))+\int_0^{t_1} \frac{{\rm d}}{{\rm d} t}(V_2^l(z_2(\tau))){\rm d}\tau\notag\\
&\leq V_2^l(z_2(0))-\int_0^{t_1} \chi_2 {\rm d} \tau\notag\\
&\leq V_2^l(z_2(0))- \chi_2  t_1,
\end{flalign}
where $l \in \{m,M\}$. An upper bound in the second dimension   of the convergence time is
\begin{equation}
T_b=\frac{1}{\chi_2}{\rm max}\{V_2^M(z_2(0)),V_2^m(z_2(0))\}.
\end{equation}

Similarly, the upper bound in the first dimension is $T_a=\frac{1}{\chi_1}({\rm max}\{V_1^M(z_1(t_1)),V_1^m(z_1(t_1))\})$.  Since $T_s=T_a+T_b$, $T_s$ can be obtained as
\begin{equation*}
\begin{split}
T_s=&\frac{1}{\chi_1}{\rm max}\{V_1^M(z_1(t_1)),V_1^m(z_1(t_1))\}\\
&+\frac{1}{\chi_2}{\rm max}\{V_2^M(z_2(0)),V_2^m(z_2(0))\}.
\end{split}
\end{equation*}
This completes the proof.
$\hfill \blacksquare$

{
\subsection{Decentralized Controller on the CPL against Byzantine Node Attacks}
Define the following state tracking error:
\begin{equation}\label{eq19}
\varepsilon_i=x_i-\Pi_i z_i,
\end{equation}
and define its statelike error:
\begin{equation}\label{eq20}
\bar{\varepsilon}_i=B_i^T P_i \varepsilon_i.
\end{equation}

The $\bar{\varepsilon}_{ij}$ denotes the $j$th element of $\bar{\varepsilon}_i$, and  $\rm{diag}(\frac{\bar{\varepsilon}_{ij}}{\sqrt{\bar{\varepsilon}_{ij}^2+\omega^2}})$ denotes a diagonal matrix with a diagonal element of $\frac{\bar{\varepsilon}_{ij}}{\sqrt{\bar{\varepsilon}_{ij}^2+\omega^2}}$. The $\hat{\kappa}_i$ denotes the estimate of $\kappa_i$, and $\tilde{\kappa}_i=\kappa_i-\hat{\kappa}_i$.

We then present the following control protocols:
\begin{align}
u_i&=G_i(\iota_i)\varrho_i\label{eq60}\\
\label{eq61}
\dot{\iota}_i&=-d_i\bar{\varepsilon}_i^T\varrho_i\\
\varrho_i&=K_ix_i+H_i z_i-\hat{\psi}_i, \label{eq21}
\end{align}
where $d_i$ is a positive constant and $\omega=-{\rm exp}(-\alpha_\omega t)$ with $\alpha_\omega$ being a positive constant. The
 $G_i(\iota)$ denotes a diagonal matrix with the main diagonal
being Nussbaum functions. In this paper, $G_i(\iota_i)$ is selected as $G_i(\iota_i)=-\alpha_i {\rm exp}(\frac{\iota_i^2}{2})(\iota_i^2+2){\rm sin}(\iota_i)$ with $\alpha_i$ being a positive const.
 The adaptive compensational signal is designed as follows
\begin{align}
\hat{\psi}_i&={\rm diag}(\frac{\bar{\varepsilon}_{ij}}{|{\bar{\varepsilon}_{ij}|+\omega}})\hat{\kappa}_i,  \label{eq22}\\
\dot{\hat{\kappa}}_i&={\rm diag}(\frac{\bar{\varepsilon}_{ij}}{{|\bar{\varepsilon}_{ij}|+\omega}}) \Bar{\varepsilon_i}, \label{eq23}
\end{align}
where ${\hat{\kappa}}_i$ is an adaptive updating parameter.

\begin{myTheo}
Consider the heterogeneous MASs with \eqref{eq6} and \eqref{eq7}.
 \textbf{Problem RC3HPB} can be solved  via the secure TL in (\ref{eq16})  and the decentralized controller (\ref{eq60})-(\ref{eq23}), if the following  condition holds simultaneously:

 \begin{enumerate}
\item The controller gain matrices $K_i$ and $H_i$ are designed as:
\begin{align}
K_i&=-U_i^{-1}B_i^T P_i, \label{eq24} \\
H_i&=\Gamma_i-K_i \Pi_i, \label{eq25}
\end{align}
where $P_i$ is a positive definite matrix and satisfies the following Riccati equation:
\begin{equation}\label{eq26}
P_i A_i+A_i^T P_i-P_i B_i U_i^{-1}B_i^T P_i +Q_i= 0,
\end{equation}
where $U_i > 0$ and $Q_i >0$ are the symmetric matrixes.
\end{enumerate}
\end{myTheo}

\textbf{Proof.} Note that \eqref{eq9} can be written as
\begin{equation*}
\begin{split}
\mathrm{e}_i(t)&=C_ix_i(t)-Rx_0(t)\\
&=C_ix_i(t)-C_i\Pi_ix_0(t)\\
&=C_i(x_i(t)-\Pi_ix_0(t))\\
&=C_i\big(x_i(t)-\Pi_i z_i(t)+\Pi_i(z_i(t)-x_0(t))\big)\\
&=C_i\varepsilon_i(t)+C_i\Pi_i(z_i(t)-x_0(t)),
\end{split}
\end{equation*}

Then we have
\begin{equation*}
\begin{split}
\Vert \mathrm{e}_i(t) \Vert& = \Vert C_i \varepsilon_i(t)+C_i\Pi_i(z_i(t)-x_0(t)) \Vert\\
&\leq \Vert C_i\varepsilon_i(t) \Vert+\Vert C_i\Pi_i(z_i(t)-x_0(t)) \Vert\\
&\leq \Vert C_i\Vert \Vert \varepsilon_i(t) \Vert+\Vert C_i\Pi_i\Vert \Vert (z_i(t)-x_0(t)) \Vert.
\end{split}
\end{equation*}

To show that $ \mathrm{e}_i(t)$ converges to 0, in the following part, we shall prove that $\Vert \varepsilon_i(t) \Vert \rightarrow 0$ and $\Vert (z_i(t)-x_0(t)) \Vert  \rightarrow 0$ as $t \rightarrow \infty$.

According to Theorem $\ref{theorem_2}$, $\Vert z_i(t)-x_0(t) \Vert$ converges to 0 asymptotically.

Next, we prove that $\Vert \varepsilon_i(t) \Vert$ converges to 0. From \eqref{eq6}, \eqref{eq7}, \eqref{eq8},  \eqref{eq16} and \eqref{eq25}, we obtain the time derivative of \eqref{eq19} as
\begin{equation*}
\begin{split}
\dot{\varepsilon}_i =&\dot{x}_i-\Pi_i (S z_i-c\beta_i)\\
=&A_i x_i+B_i K_i x_i+B_i H_i z_i+B_i (\psi_i-\hat{\psi}_i)
-\Pi_i S z_i\\
&+c\Pi_i \beta_i+B_i(m_i(t)G_i(\iota_i)-1)\varrho_i\\
=&\bar{A}_i \varepsilon_i+B_i (\psi_i- \hat{\psi}_i)+c \Pi_i \beta_i+B_i(m_i(t)G_i(\iota_i)-1)\varrho_i,
\end{split}
\end{equation*}
where $\bar{A}_i = A_i+B_i K_i$ and $\beta_i=[\sum_{j \in \mathcal{N}_i}{\rm sign}(z_{i,1}-z_{j,1});\sum_{j \in \mathcal{N}_i}{\rm sign}(z_{i,2}-z_{j,2})]$.

Let $\bar{Q}_i=Q_i+K_i^T U_i K_i$. It can be obtained from \eqref{eq26}  that
\begin{equation}
P_i \bar{A}_i+\bar{A}_i^T P_i= -\bar{Q}_i.
\end{equation}

The Lyapunov function is considered as:
\begin{equation}\label{34}
V(t)= \varepsilon_i^T P_i \varepsilon_i+\tilde{\kappa}_i^T \tilde{\kappa}_i.
\end{equation}

The derivative of $V(t)$ can be calculated as
\begin{flalign}\label{eq35}
\dot{V}(t)=&2 \varepsilon_i^T P_i\dot{\varepsilon}_i-2\tilde{\kappa}_i^T \dot{\hat{\kappa}}_i\notag&\\
=&2 \varepsilon_i^T P_i(\bar{A}_i \varepsilon_i-B_i \hat{\psi}_i+B_i \psi_i+c\Pi_i \beta_i)\notag&\\
&+2\bar{\varepsilon}_i^T(m_i(t)G_i(\iota_i)-1)\varrho_i-2\tilde{\kappa}_i^T \dot{\hat{\kappa}}_i\notag&\\
=& \varepsilon_i^T(P_i \bar{A}_i+\bar{A}_i^T P_i) \varepsilon_i+2c~ \varepsilon_i^T P_i \Pi_i \beta_i\notag&\\
&+2 \varepsilon_i^T P_i (B_i \psi_i-B_i \hat{\psi}_i)\notag&\\
&+2\bar{\varepsilon}_i^T(m_i(t)G_i(\iota_i)-1)\varrho_i-2\tilde{\kappa}_i^T \dot{\hat{\kappa}}_i\notag&\\
=& -\varepsilon_i^T\bar{Q}_i\varepsilon_i+2 \varepsilon_i^T P_i (B_i \psi_i-B_i \hat{\psi}_i)\notag&\\
&+2\bar{\varepsilon}_i^T(m_i(t)G_i(\iota_i)-1)\varrho_i-2\tilde{\kappa}_i^T \dot{\hat{\kappa}}_i.&
\end{flalign}

Invoking Theorem 2, we conclude that all agents can track
the leader in finite time. Then, the $z_i-z_j=0$ holds in finite time, which means the sign function in \eqref{eq16} will equal to zero.

By using \eqref{eq23} and $|\bar{\varepsilon}_{ij}| \kappa_{ij}-\bar{\varepsilon}_{ij} (\frac{\bar{\varepsilon}_{ij}}{|\bar{\varepsilon}_{ij}|+\omega}) \kappa_{ij}\leq \omega \kappa_{ij} \leq \omega \kappa_{\rm max}$, we can obtain
\begin{align}\label{eq36}
&2\varepsilon_i^T P_i B_i (\psi_i-\hat{\psi}_i)-2\tilde{\kappa}_i^T \dot{\hat{\kappa}}_i\notag\\
\leq&2|\bar{\varepsilon}_i|^T \kappa_i-2\bar{\varepsilon}_i^T{\rm diag}(\frac{\bar{\varepsilon}_{ij}}{|\bar{\varepsilon}_{ij}|+\omega})\hat{\kappa}_i\notag\\
&-2(\kappa_i^T-\hat{\kappa}_i^T){\rm diag}(\frac{\bar{\varepsilon}_{ij}}{|\bar{\varepsilon}_{ij}|+\omega}) \Bar{\varepsilon_i}\notag\\
\leq&2|\bar{\varepsilon}_i|^T \kappa_i-2\bar{\varepsilon}_i^T {\rm diag}(\frac{\bar{\varepsilon}_{ij}}{|\bar{\varepsilon}_{ij}|+\omega}) \kappa_i\notag\\
\leq&2\omega \kappa_{\rm max}.
\end{align}
where $\kappa_i$ is defined in Assumption \ref{assumption_6}. Substituting $\eqref{eq61}$ and  $\eqref{eq36}$ into $\eqref{eq35}$ yields,
\begin{equation}\label{eq39}
\dot{V}(t)\leq -\varepsilon_i^T\bar{Q}_i\varepsilon_i
-\frac{2}{d_i}(m_i(t)G_i(\iota_i)-1)\dot{\iota}_i+2\omega \kappa_{\rm max}.
\end{equation}
Then, we have
\begin{equation}\label{eq44}
\begin{split}
\dot{V}(t) \leq -\eta  \varepsilon_i^T \varepsilon_i(t) -\frac{2}{d_i}(m_i(t)G_i(\iota_i)-1)\dot{\iota}_i+2\omega \kappa_{\rm max},
\end{split}
\end{equation}
where $\eta = \sigma_{\mathrm{min}}(\bar{Q}_i)$. From \eqref{eq44}, we have
\begin{align}\label{eq45}
V(t)-V(0)\leq -\eta  \int_{0}^t \Vert \varepsilon_i(\tau) \Vert^2 {\rm d} \tau -\frac{2}{d_i}E_1+E_2,
\end{align}
where $E_1=\int_{0}^t (m_i(\tau) G_i(\iota_i)-1)\dot{\iota}_i {\rm d}\tau$ and $E_2=\int_{0}^t 2\omega
\kappa_{\rm max} \\ {\rm d}\tau$. The $E_2$ in $\eqref{eq45}$ is bounded. In addition,  the boundedness of $E_1$ can be achieved by seeking a suitable function as shown in \cite{chen2016adaptive}-\cite{wang2016adaptive}.

Thus, we have
\begin{equation}\label{eq46}
 \Vert \varepsilon_i(t) \Vert^2 \leq -  \int_{0}^t \frac{\eta}{\sigma_{\mathrm{min}}({P}_i)}  \Vert \varepsilon_i(\tau) \Vert^2 {\rm d} \tau + E,
\end{equation}
where $E=\frac{1}{\sigma_{\mathrm{min}}({P}_i)} (V(0)+{\rm sup}_{t \geq 0}(-\frac{2}{d_i} |E_1|+|E_2 |))$.
Using Bellman–Gronwall Lemma, we obtain
\begin{equation}\label{eq47}
 \Vert \varepsilon_i(t) \Vert \leq \sqrt{E}\mathrm{e}^{-\frac{\eta}{2\sigma_{\mathrm{min}}({P}_i)} t}.
\end{equation}
That is, $\Vert \varepsilon_i(t) \Vert$ exponentially converges to zero.
Hence, the cooperative control problem for MASs \eqref{eq6} has been solved, which completes the proof.  $\hfill \blacksquare$

\begin{remark}
{
The main merits of this work are threefold:
\begin{enumerate}
  \item Compared with the recent work \cite{gusrialdi2018competitive}, this paper studies the more complex heterogeneous dynamics.

  \item Different from the existing work \cite{2021ResilientY} only ensuring the performance of normal agents, we manage to ensure both Byzantine agents and normal agents
  achieve the consensus
  in the presence of Byzantine attacks via our twin layer approach. Thus, we first achieve resilient control \textbf{against} Byzantine attacks, rather than \textbf{under} Byzantine attacks.
$\hfill \hfill \square $

\end{enumerate}
}
\end{remark}

\section{Numerical Simulation}
In this section, we employ a simulation example to verify the effectiveness of the above theoretical results.

Our goal is to achieve resilient cooperative  control of heterogeneous MASs. Consider heterogeneous MASs of six agents. Their communication
topology is shown in Figure. \ref{image2}. The heterogeneous dynamics of (\ref{eq6})  can be described as $\alpha_{s_i}=-1, \alpha_{v_i}=-2, i\in\textbf{I}[1,3]$ and $\alpha_{s_i}=-2, \alpha_{v_i}=-3, i\in\textbf{I}[4,6]$.




We consider the following kinds of Byzantine attacks, that is, agents 1 and 5 are Byzantine agents which satisfy $f$-local assumption, that is, Assumption \ref{assumption_5}. The Byzantine edge attacks on the TL  are to replace the information among agents as:
\begin{equation*}
e_{1,5}= \left[  \begin{array}{ccc}
t  \\
2t   \\
  \end{array}\right], e_{1,6}=  \left[  \begin{array}{ccc}
1.5t   \\
-2.5t  \\
  \end{array}\right],
  e_{5,4}=  \left[  \begin{array}{ccc}
-1.5t   \\
2.5t   \\
  \end{array}\right].
\end{equation*}
where $e_{i,j}$ denotes the information flow from agent $i$ to agent $j$ falsified by the Byzantine edge attacks.
Notice that $e_{1,5}(t)$, $e_{1,6}(t)$ and $e_{5,4}(t)$ are different from each other, which illustrates the difference between Byzantine attacks  and other malicious attacks. The Byzantine node attacks are
\begin{equation*}
\psi_3= 2{\rm sin}(0.5t),  ~ \psi_5=  -2{\rm sin}(0.5t),
\end{equation*}
and
\begin{equation*}
m_i(t)=-0.6+0.1{\rm cos}(t)
\end{equation*}

We conduct the resilient design of the TL. The detailed construction process of TL is summarized as Example 1 in Section \ref{IV_A}.
The distributed estimation performance of the TL
against Byzantine edge attack is shown in Fig. \ref{image4} and Fig. \ref{image5}. It is shown that the asymptotic performance of distributed estimation
error on the TL can be achieved despite the Byzantine edge attacks.

Then, we focus on the performance of CPL with the consideration of the potential  Byzantine node attacks.
In light of \eqref{eq47}, the Byzantine node attack signals satisfy Assumption 6. It can be computed from \eqref{eq8} that
\begin{align*}
&\Pi_{1,2,3}=  \left[  \begin{array}{ccc}
1 &  0  \\
0 &  1  \\
  \end{array}\right], ~\Gamma_{1,2,3}= \left[  \begin{array}{ccc}
1 & 2   \\
  \end{array}\right],\\
&\Pi_{4,5,6}=\left[  \begin{array}{ccc}
1 &  0  \\
0 &  1   \\
  \end{array}\right],~ \Gamma_{4,5,6}=  \left[  \begin{array}{ccc}
2 & 3   \\
  \end{array}\right].
\end{align*}

By selecting $U_{1\sim6}=I_2$ and $w=1$, it can be obtained that
\begin{align*}
    &P_{1,2,3}=  \left[  \begin{array}{ccc}
4 & 1\\
1 &  1  \\
  \end{array}\right], ~P_{4,5,6}=  \left[  \begin{array}{ccc}
4.90 & 0.90   \\
0.90 & 0.90  \\
  \end{array}\right],\\
&K_{1,2,3}=  \left[  \begin{array}{ccc}
-1 &  -1  \\
  \end{array}\right], ~H_{1,2,3}=  \left[  \begin{array}{ccc}
2 & 3  \\
  \end{array}\right],\\
&K_{4,5,6}=  \left[  \begin{array}{ccc}
-0.45 &  -0.45  \\
  \end{array}\right], ~H_{4,5,6}=  \left[  \begin{array}{ccc}
2.45 & 3.45  \\
  \end{array}\right].
\end{align*}

%
%

By employing the above parameters, the cooperative control error is recorded in Fig. \ref{image6}. The exact agent error $\varepsilon_i$ between the CPL and the TL  is depicted in Fig. \ref{image7}, which is proved to converge to zero.

\section{Conclusion}
In this paper, we investigate the problem of cooperative control of heterogeneous MASs in the presence of Byzantine attacks. Apart from CPL, a virtual TL is deployed on the cloud. The double-layer  scheme decouples the defense strategy against Byzantine attacks into the defense against Byzantine edge attacks on the TL and the defense against Byzantine node attacks on the CPL. On the TL, we propose a resilient topology reconfiguration strategy by adding a minimum number of key edges to ensure the $(2f+1)$-robustness of network topology.  A  local interaction protocol based on the sign function  works resiliently against  Byzantine edge attacks.  All the virtual state can follow the leader.
On the CPL, a decentralized controller is designed against  Byzantine node attacks, which ensures exponential convergence. Based on the theoretical result,
the agent is achieved against Byzantine attacks.
In future work, it is interesting to consider the consensus problem of MASs with nonlinear dynamics \cite{2019Leader}.

}


 \bibliography{20230109}
 \end{document}